\documentclass{aa}

\usepackage{graphicx}
\usepackage{epsfig}

\newcommand{\pl}{\partial}
\renewcommand{\d}{{\rm d}}
\newcommand{\inta}{\int_{-i\infty}^{+i\infty}}
\newcommand{\beq}{\begin{equation}}
\newcommand{\eeq}{\end{equation}}
\newcommand{\beqa}{\begin{eqnarray}}
\newcommand{\eeqa}{\end{eqnarray}}
\newcommand{\bea}{\begin{array}}
\newcommand{\ea}{\end{array}}
\newcommand{\bx}{{\bf x}}
\newcommand{\bp}{{\bf p}}
\newcommand{\cG}{{\cal G}}
\newcommand{\df}{\delta f}
\newcommand{\rhob}{\overline{\rho}}
\newcommand{\bk}{{\bf k}}
\newcommand{\lag}{\langle}
\newcommand{\rag}{\rangle}
\newcommand{\bv}{{\bf v}}
\newcommand{\Om}{\Omega_{\rm m}}

\newcommand{\cO}{{\cal O}}
\newcommand{\cOm}{{\cal O}_{-1}}
\newcommand{\br}{{\bf r}}
\newcommand{\tK}{{\tilde K}}
\newcommand{\dfn}{\delta f^{(n)}}
\newcommand{\tOm}{{\tilde {\cal O}}_{-1}}
\newcommand{\tKs}{{\tilde K}_s}
\newcommand{\dLo}{\delta_{L0}}
\newcommand{\cK}{{\cal K}}
\newcommand{\bom}{{\vec \omega}}
\newcommand{\dg}{{\dot g}}
\newcommand{\cNo}{{\cal N}_0}
\newcommand{\Dem}{\Delta_{\eta}^{-1}}
\newcommand{\De}{\Delta_{\eta}}
\newcommand{\Det}{\mbox{Det}}
\newcommand{\Tr}{\mbox{Tr}}
\renewcommand{\Re}{\mbox{Re}}

\begin{document}

 

\title{Dynamics of gravitational clustering I. Building perturbative expansions.}   
\author{P. Valageas}  
\institute{Service de Physique Th\'eorique, CEN Saclay, 91191 Gif-sur-Yvette, France} 
\date{Received / Accepted }

\abstract{
We develop a systematic method to obtain the solution of the {\it collisionless Boltzmann equation} which describes the growth of large-scale structures as a perturbative series over the initial density perturbations. We give an explicit calculation of the second-order terms which are shown to agree with the results obtained from the hydrodynamical description of the system. Then, we explain that this identity extends to all orders of perturbation theory and that the perturbative series actually diverge for hierarchical scenarios. However, since the collisionless Boltzmann equation provides the exact description of the dynamics (including the non-linear regime) these results may serve as a basis for a study of the non-linear regime. In particular, we derive a non-perturbative quadratic integral equation which explicitly relates the actual non-linear distribution function to the initial conditions (more precisely, to the linear growing mode). This allows us to write an explicit path-integral expression for the probability distribution of the exact non-linear density field.
}

\maketitle

\section{Introduction}

In usual cosmological models large-scale structures in the universe have formed through the growth of small initial density perturbations by gravitational instability, see \cite{Peebles1}. Moreover, in most cases of cosmological interest the power increases at small scales as in the CDM model (\cite{Peebles2}). This leads to a hierarchical scenario of structure formation where smaller scales become non-linear first. Once they enter the non-linear regime high-density fluctuations form massive objects which will give birth to galaxies or clusters. Thus, it is important to understand the dynamics of these density perturbations in order to describe the large-scale structures we observe in the present universe.

The distribution of matter is usually described as a pressure-less fluid through the standard equations of hydrodynamics (continuity and Euler eq.) coupled to the Poisson equation for the gravitational potential. Then, one often looks for a solution of the equations of motion under the form of a perturbative expansion (e.g., \cite{Fry1}, \cite{Goroff1}, \cite{Ber2}). In fact, one cannot hope to describe the non-linear regime within such a framework since this hydrodynamical description becomes inexact as soon as shell-crossing appears. Then, there is no longer only one velocity at each point. Note that for hierarchical scenarios with no small-scale cutoff (i.e. the amplitude of density fluctuations diverges at small scales $R \rightarrow 0$) shell-crossing occurs as soon as $t>0$. Moreover, in the non-linear regime shell-crossing should play a key role (e.g., through virialization processes). 

Therefore, in order to obtain a complete description of the dynamics one needs to go beyond the standard approximation of a pressure-less fluid. Then, one might add a pressure term to the hydrodynamical equations in order to describe the effects of multi-streaming, which provide some velocity dispersion (e.g., \cite{Dom1}, \cite{Dom2}). However, this is not very satisfactory because collisionless and gaseous systems can exhibit very different behaviours starting from the same initial conditions. For instance, some physical processes are specific to collisionless systems (e.g., Landau damping, phase-mixing) and stability conditions for similar collisionless and gaseous systems are not always the same, see \cite{Bin1}. Hence it is important to investigate the dynamics of density fluctuations within the framework of the collisionless Boltzmann equation, which provides an exact description of the system both in the linear and non-linear regimes.

Since this is a rather difficult task we may first try to devise a perturbative approach. This may then serve as a basis for further developments to tackle the non-linear regime. Thus, in this article we build a perturbative approach to the collisionless Boltzmann equation. One could also start from the BBGKY hierarchy, which can be derived from the Boltzmann equation. Such a study was performed in \cite{Bha1} where it was shown that the lowest non-linear order matches the results of the usual hydrodynamical approach. We recover this result here and we explain that indeed a perturbative method cannot handle shell-crossing. However, in the building of perturbative theory we also derive in this article some non-perturbative results which may prove to be useful for a study of the non-linear regime.

This paper is organized as follows. In Sect.\ref{Equations of motion} we write the equations of motion which describe the dynamics of the collisionless fluid. Then, in Sect.\ref{Building a perturbative expansion} we develop a systematic procedure to obtain the fluctuations of the distribution function as a perturbative series over the initial perturbations (or more precisely over the linear growing mode) in the case of a critical density universe. In doing so we also derive a non-perturbative quadratic integral equation which directly relates the actual non-linear distribution function to this linear mode. Next, we give in Sect.\ref{Explicit calculations} the explicit calculations of the second-order terms. We explain in Sect.\ref{Divergence of perturbative series} why we recover the results of the hydrodynamical approach and why the perturbative series must diverge for hierarchical scenarios (with no small-scale cutoff). Then, we show in Sect.\ref{Other cosmologies} how to extend this method to arbitrary cosmologies. Finally, we derive in Sect.\ref{Functional formulation} an explicit path-integral expression which describes the statistical properties of the exact non-linear density field.

\section{Equations of motion}
\label{Equations of motion}

The gravitational dynamics of a fluid of collisionless particles of mass $m$ is given by the collisionless Boltzmann equation (\cite{Peebles1}), coupled with the Poisson equation:
\beq
\left\{ \bea{l} {\displaystyle
 \frac{\pl f}{\pl t} + \frac{\bp}{m a^2} . \frac{\pl f}{\pl \bx} - m \frac{\pl \phi}{\pl \bx} . \frac{\pl f}{\pl \bp} = 0 }  \\ \\ {\displaystyle
  \Delta \phi = \frac{4\pi \cG}{a} (\rho - \rhob) \hspace{0.2cm} , \hspace{0.2cm} \rho(\bx,t) = m \int f(\bx,\bp,t) \; \d\bp }  \ea \right.
\label{Bol1}
\eeq
where $f(\bx,\bp,t)$ is the distribution function (phase-space density), $\phi$ is the gravitational potential, $\rho(\bx,t)$ is the comoving density, $\rhob$ is the mean comoving density of the universe (hence it is independent of time), $a(t)$ is the scale factor and the momentum $\bp$ is related to the comoving coordinate $\bx$ of a particle by $\bp = m a^2 \dot{\bx}$. Thus, eq.(\ref{Bol1}) describes the growth of gravitational structures in an expanding universe which is homogeneous on large scales. 

Note that in order to derive the collisionless Boltzmann equation one needs to neglect the two-particle correlations. Then, the gravitational interaction is described by a smooth gravitational potential which neglects the effects of the discrete character of the distribution of matter (e.g., \cite{Peebles1}). This becomes exact in the continuous limit $m \rightarrow 0$ which we perform below in eq.(\ref{Bol2}). However, one must also add some small-scale cutoff to the power-spectrum of the density fluctuations in order to guarantee small-scale smoothness.

The definition itself of the problem we investigate here also involves an implicit ``regularization'' of the gravitational interaction. Indeed, within an infinite uniform universe the gravitational force is not well defined since the pull due to the matter located within a small solid angle $\d\Omega$ diverges as we integrate over all shells up to infinity. This problem is cured by first integrating the relevant integrals over angles, and next over radial distances (e.g., \cite{Peebles1}). This can also be seen as the introduction of a large-scale cutoff $L$ for the gravitational interaction which preserves rotational symmetry. More precisely, this means that in Fourier space we integrate on the wavenumber $\bk$ over $|\bk|>k_c$ and in the final results we take the limit $k_c \rightarrow 0$. In fact, except in eq.(\ref{T2}) in Sect.\ref{Functional formulation} this feature will not show up in the equations we encounter in this article so that we can directly take $k_c=0$. 

Next, we can absorb the mass $m$ into the distribution function $f$ and the momentum $\bp$: $f \rightarrow f/m^4$, $\bp \rightarrow m \bp$, and we obtain:
\beq
\left\{ \bea{l} {\displaystyle
 \frac{\pl f}{\pl t} + \frac{\bp}{a^2} . \frac{\pl f}{\pl \bx} - \frac{\pl \phi}{\pl \bx} . \frac{\pl f}{\pl \bp} = 0 }  \\ \\ {\displaystyle
  \Delta \phi = \frac{4\pi \cG}{a} (\rho - \rhob) \hspace{0.2cm} , \hspace{0.2cm} \rho(\bx,t) = \int f(\bx,\bp,t) \; \d\bp }  \ea \right.
\label{Bol2}
\eeq
which is also valid in the continuous limit $m \rightarrow 0$. Here $f(\bx,\bp,t) \d^3x \d^3p$ is the mass enclosed in the phase-space element $\d^3x \d^3p$ and the momentum $\bp$ verifies:
\beq
\bp = a^2 \dot{\bx} .
\label{p1}
\eeq
In order to single out the deviations of the distribution function from the case of a perfectly homogeneous universe we define the perturbation $\df(\bx,\bp,t)$ by:
\beq
f(\bx,\bp,t) \equiv \rhob \; \delta_D(\bp) + \rhob \; \df(\bx,\bp,t)
\label{df1}
\eeq
where $\delta_D$ is Dirac's function (here in 3 dimensions). Then, the density contrast $\delta(\bx,t)$ is simply given by:
\beq
\delta(\bx,t) = \int \df(\bx,\bp,t) \; \d\bp .
\label{delta1}
\eeq
From the definition (\ref{df1}) and the equation of motion (\ref{Bol2}) we obtain the evolution equation for $\df$:
\beq
\left\{ \bea{l} {\displaystyle
 \frac{\pl \df}{\pl t} + \frac{\bp}{a^2} . \frac{\pl \df}{\pl \bx} - \frac{\pl \phi}{\pl \bx} . \frac{\pl}{\pl \bp} \delta_D(\bp) - \frac{\pl \phi}{\pl \bx} . \frac{\pl \df}{\pl \bp} = 0 }  \\ \\ {\displaystyle
  \Delta \phi = \frac{4\pi \cG}{a} \rhob \; \delta(\bx,t) }  \ea \right.
\label{Bol3}
\eeq
Note the term $\delta_D'(\bp)$ which is sometimes forgotten in the literature. Next, we define the spatial Fourier transform of the field $\df$ by:
\beq
\left\{ \bea{l} {\displaystyle \df(\bx,\bp,t) = \int \d\bk \; e^{i \bk.\bx} \; \df(\bk,\bp,t) }  \\ \\ {\displaystyle \df(\bk,\bp,t) = \frac{1}{(2\pi)^3} \int \d\bx \; e^{-i \bk.\bx} \; \df(\bx,\bp,t) } \ea \right.
\label{Four1}
\eeq
To simplify the notations, we note the distribution function by $\df$, both in real space $\bx$ and in Fourier space $\bk$. Its argument removes any ambiguity. We also define the Fourier transforms $\delta(\bk)$ and $\phi(\bk)$ of the density contrast and of the gravitational potential as in eq.(\ref{Four1}). Then, Poisson's equation reads:
\beq
-k^2 \phi(\bk) = \frac{4\pi \cG}{a} \rhob \; \delta(\bk) .
\label{Poisson}
\eeq
After substitution of eq.(\ref{Poisson}) into the equation of motion (\ref{Bol3}) we obtain:
\beqa
\frac{\pl \df}{\pl t} + i \frac{\bk.\bp}{a^2} \df + i \frac{4\pi \cG \rhob}{a}  \frac{\bk}{k^2} . \frac{\pl \delta_D}{\pl \bp}(\bp) \int \d\bp' \; \df(\bk,\bp') \nonumber \\  \hspace{0.cm} + i \frac{4\pi \cG \rhob}{a} \int \d\bk' \d\bp' \df(\bk',\bp') \frac{\bk'}{k'^2} . \frac{\pl \df}{\pl \bp}(\bk-\bk',\bp) = 0 .
\label{Bol4}
\eeqa
This {\it non-linear} and {\it non-local} equation describes the dynamics of the collisionless fluid.

\section{Building a perturbative expansion}
\label{Building a perturbative expansion}

\subsection{General framework}
\label{General framework}

Here we consider the case of a critical density universe $\Om=1$. Then, the average comoving density is given by:
\beq
\rhob = \frac{a^3}{6 \pi \cG t^2} .
\eeq
We normalize the time scale such that the Hubble time today $t_0$ (i.e. at $z=0$) satisfies $t_0=1$, while the scale factor is $a=1$. Then, we define a new time coordinate $\tau$ by:
\beq
\tau \equiv \frac{1}{3}\ln (t/t_0) = \frac{1}{3}\ln t , \hspace{0.2cm} t=e^{3\tau} , \hspace{0.2cm} a = e^{2\tau}
\label{tau}
\eeq
and eq.(\ref{Bol4}) reads:
\beqa
\lefteqn{ \frac{\pl \df}{\pl \tau} + 3 i (\bk.\bp) e^{-\tau} \df + 2 i \; e^{\tau} \frac{\bk}{k^2} . \frac{\pl \delta_D}{\pl \bp}(\bp) \int \d\bp' \; \df(\bk,\bp')} \nonumber \\ & & + 2 i \; e^{\tau} \int \d\bk' \d\bp' \df(\bk',\bp') \frac{\bk'}{k'^2} . \frac{\pl \df}{\pl \bp}(\bk-\bk',\bp) = 0 .
\label{Bol5}
\eeqa
This equation is quadratic over the distribution function $\df$. In order to make apparent the structure of this equation of motion it is convenient to separate the linear and quadratic terms. Thus, we can also write eq.(\ref{Bol5}) as:
\beq
(\cO . \df) (\br) = g \int \d^7r_1 \d^7r_2 \; K(\br;\br_1,\br_2) \; \df(\br_1) \; \df(\br_2)
\label{Bol6}
\eeq
where we note $\br$ the 7-dimensional coordinate $\br=(\bk,\bp,\tau)$ and we introduced the linear operator $\cO$ and the kernel $K$ defined by:
\beqa
\lefteqn{ (\cO . \df) (\br) \equiv \frac{\pl \df}{\pl \tau} + 3 i (\bk.\bp) \; e^{-\tau} \; \df } \nonumber \\ & & + 2 i \; e^{\tau} \; \frac{\bk}{k^2} . \frac{\pl \delta_D}{\pl \bp}(\bp) \int \d\bp' \; \df(\bk,\bp',\tau)
\label{O1}
\eeqa
and:
\beqa
K(\br;\br_1,\br_2) & \equiv & 2 i \; e^{\tau} \; \delta_D(\tau_1-\tau) \delta_D(\tau_2-\tau) \nonumber \\  & & \times \delta_D(\bk_1+\bk_2-\bk) \; \frac{\bk_1}{k_1^2} . \frac{\pl \delta_D}{\pl \bp}(\bp_2-\bp) .
\label{K1}
\eeqa
In the r.h.s. in eq.(\ref{Bol6}) the integration over $\br_1$ and $\br_2$ runs over all space. In particular, the time coordinates $\tau_1$ and $\tau_2$ are integrated over the whole range $]-\infty,+\infty[$. Note however that the Dirac factor $\delta_D(\tau_1-\tau) \delta_D(\tau_2-\tau)$ which appears in the kernel $K$ implies that the r.h.s. only depends on the distribution function $\df$ at the time $\tau$, as in eq.(\ref{Bol5}). This also ensures that causality is not violated. In the relation (\ref{Bol6}) we also introduced a ``coupling constant'' $g$ which is equal to unity: $g=1$. The advantage of generalizing the quadratic equation (\ref{Bol6}) to arbitrary $g$ will appear later on.

\subsection{Perturbative expansion}
\label{Perturbative expansion}

The dynamics of the distribution function $\df$ is fully determined by the non-linear equation (\ref{Bol6}) supplemented by initial conditions. At early times $\tau \rightarrow -\infty$ the universe becomes homogeneous and the velocity is given by the Hubble flow (i.e. $\bp \rightarrow 0$) so that the perturbations vanish: $\df \rightarrow 0$. Thus, at early times (or at large scales for CDM-like initial conditions) the equations of motion can be linearized and the distribution function is equal to the linear solution $\eta(\br)$. As time goes on, the perturbation grows and it eventually enters the non-linear regime. In the mildly non-linear regime, one can seek a solution of eq.(\ref{Bol6}) in the form of a perturbative expansion over the linear mode $\eta(\br)$. Thus, we write the expansion:
\beq
\df(\br) = \sum_{n=1}^{\infty} \df^{(n)}(\br)
\label{exp1}
\eeq
with:
\beq
\df^{(1)}(\br) \equiv \eta(\br) \hspace{0.2cm} \mbox{and} \hspace{0.2cm} \df^{(n)}(\br) \sim \eta^n .
\label{exp2}
\eeq
That is, the term of order $n$ is of the order of $\eta$ to the power $n$. Substituting the expansion (\ref{exp1}) into the equation of motion (\ref{Bol6}) we obtain:
\beq
\cO . \eta = 0
\label{ker}
\eeq
and for $n \geq 2$:
\beq
\cO.\df^{(n)} = g \int \d\br_1 \d\br_2 \; K \; \sum_{m=1}^{n-1} \df^{(m)}(\br_1) \df^{(n-m)}(\br_2) .
\label{rec1}
\eeq
Thus, the linear mode $\eta(\br)$ must belong to the kernel of the operator $\cO$. Since we require $\eta \neq 0$ this implies that $\cO$ is not invertible. We shall check below that indeed the kernel of the operator $\cO$ defined by eq.(\ref{O1}) is not reduced to $\{0\}$. Then, the higher-order terms $\df^{(n)}$ are given by the recursion (\ref{rec1}). 

However, in order to obtain an explicit expression for $\df^{(n)}$ we must ``invert'' the operator $\cO$ in eq.(\ref{rec1}). More precisely, we need to build a kernel $\tK(\br;\br_1,\br_2)$ which satisfies:
\beq
(\cO.\tK)(\br;\br_1,\br_2) = K(\br;\br_1,\br_2) .
\label{tK0}
\eeq
Indeed, if such a kernel $\tK$ exists, we can solve the recursion (\ref{rec1}) by choosing for the term of order $n$ the value:
\beq
\df^{(n)} = g \int \d\br_1 \d\br_2 \; \tK \; \sum_{m=1}^{n-1} \df^{(m)}(\br_1) \df^{(n-m)}(\br_2) .
\label{rec2}
\eeq
This relation fully defines the expansion (\ref{exp1}): it provides an explicit procedure to compute any order term $\df^{(n)}$, in a recursive fashion. Note that it is not obvious a priori that a solution $\tK$ to eq.(\ref{tK0}) exists. On the other hand, if a solution exists it is not unique since we can add to any peculiar solution $\tK$ any factorized kernel of the form $\eta(\br) H(\br_1,\br_2)$ where $H$ is arbitrary, as shown by eq.(\ref{ker}). This amounts to add a factor $\eta$ to all terms $\df^{(n)}$. We shall come back to this point below.

Thus, we now need to construct a solution $\tK$ to eq.(\ref{tK0}). To do so, we use the following trick. First, we define a new operator $\cOm$ by:
\beq
\cOm(\br) \equiv \int_0^{\infty} \d \sigma \; e^{-\sigma \cO(\br)} .
\label{Om1}
\eeq
Since the operator $\cOm$ is only an intermediate tool eq.(\ref{Om1}) should be interpreted in a formal sense. In particular, one may introduce a large cutoff $\Lambda$ in the intermediate steps of the calculation for the integration over $\sigma$ in the definition (\ref{Om1}) and take the limit $\Lambda \rightarrow \infty$ in the final results. The reason behind the definition (\ref{Om1}) is that this operator $\cOm$ is formally the ``inverse'' of the operator $\cO$. Indeed, expanding the exponential $e^{-\sigma \cO}$ we can write:
\beq
\frac{\pl}{\pl \sigma} \left( e^{-\sigma \cO} . f \right) = \sum_{q=1}^{\infty} \frac{(-1)^q \; \sigma^{q-1}}{(q-1)!} \; \cO^q . f = - \cO . e^{-\sigma \cO} . f
\label{expon1}
\eeq
where $f(\br)$ is an arbitrary function of $\br$. Using eq.(\ref{expon1}) we obtain:
\beqa
\lefteqn{\cO . \cOm . f = \int \d \sigma \, \cO . e^{-\sigma \cO} . f = - \int_0^{\infty} \d \sigma \, \frac{\pl}{\pl \sigma} \left( e^{-\sigma \cO} . f \right) = f } \nonumber \\ 
\label{expon2}
\eeqa
if all integrals converge. Here, we must point out that eq.(\ref{expon2}) is not a rigorous proof. In particular, using the same arguments we could also obtain $\cOm . \cO . f=f$. However, we know that $\cOm . \cO$ cannot be the identity operator since the kernel of $\cO$ is not reduced to $\{0\}$, as shown by eq.(\ref{ker}) which implies $\cOm . \cO . \eta=0$. Nevertheless, we can still work with the operator $\cOm$. In particular, we define a kernel $\tK(\br;\br_1,\br_2)$ by:
\beq
\tK(\br;\br_1,\br_2) \equiv \cOm(\br) . K(\br;\br_1,\br_2) .
\label{tK1}
\eeq
If the operator $\cO$ were invertible and all integrals were convergent the kernel $\tK$ would be the unique solution of eq.(\ref{tK0}). Here, the operator $\cO$ is not invertible hence it is not obvious a priori that the kernel $\tK$ defined by eq.(\ref{tK1}) is a solution of eq.(\ref{tK0}). However, if this procedure provides a finite kernel $\tK$ we could expect that it obeys eq.(\ref{tK0}). We compute in App.\ref{Calculation of the exponential} and App.\ref{Calculation of the kernel} the kernel $\tK$ defined by eq.(\ref{tK1}). This yields the equivalent expressions (\ref{tK2}) and (\ref{tK3}). As discussed in App.\ref{Calculation of the kernel} we also explicitly check that the kernel $\tK$ written in eq.(\ref{tK3}) is a solution to eq.(\ref{tK0}). Then, we obtain the explicit perturbative expansion of the distribution function $\df$ through eq.(\ref{rec2}). As noticed above, at each step $n$ we could add to $\dfn$ an element of the kernel of $\cO$ like $\eta$. However, since we require that $\dfn$ be of the order of $\eta$ to the power $n$ (so that a perturbative expansion makes sense) this additional term is zero and the high-order terms $\dfn$ are indeed given by eq.(\ref{rec2}) for $n \geq 2$.

Here we note that instead of the operator $\cOm$ defined by eq.(\ref{Om1}) we could have tried to use the operator $\tOm$ defined by:
\beq
\tOm(\br) \equiv - \int_0^{\infty} \d \sigma \; e^{\sigma \cO(\br)} .
\label{tOm1}
\eeq
Indeed, we can easily see that this operator also satisfies $\cO . \tOm = 1$ in a formal sense, following the steps (\ref{expon1}) and (\ref{expon2}). However, only one of the two choices $\cOm$ and $\tOm$ is valid and leads to finite results while the other one involves divergent quantities. In fact, it is easy to see a priori that the correct choice is $\cOm$ from causality requirements. Indeed, from eq.(\ref{O1}) we note that $\cO = \pl / \pl \tau + ...$. Then, since we have:
\beq
e^{ - \sigma \; \pl / \pl \tau} . f(\tau) \equiv \sum_{q=0}^{\infty} \frac{(-\sigma)^q}{q !} \; \frac{\pl^q}{\pl \tau^q} \; f(\tau) = f(\tau-\sigma)
\eeq
we can see that $\dfn(\tau)$ defined by eq.(\ref{rec2}) using $\cOm$ involves the linear mode at earlier times $\eta(\tau-\sigma)$ (with $\sigma \geq 0$), as required by causality. On the contrary, using $\tOm$ would imply that $\dfn(\tau)$ depend on $\eta(\tau+\sigma)$, that is on the linear mode at later times, which violates causality. As a consequence, the correct choice is the operator $\cOm$. However, the actual justification of eq.(\ref{tK1}) is the explicit check (discussed in App.\ref{Calculation of the kernel}) that the kernel $\tK$ derived in App.\ref{Calculation of the exponential} and App.\ref{Calculation of the kernel} from the definition (\ref{tK1}) is a solution to eq.(\ref{tK0}).

From the recursion (\ref{rec2}) we can also express $\dfn$ explicitly in terms of the linear mode $\eta$. Indeed, we can easily check that we can write eq.(\ref{rec2}) as:
\beq
\dfn(\br) = g^{n-1} \int \d\br_1 .. \d\br_n \; F_n(\br;\br_1,..,\br_n) \; \eta(\br_1) .. \eta(\br_n)
\label{Fn1}
\eeq
for $n \geq 1$. Here we defined the kernels $F_n$ by:
\beq
F_1(\br;\br_1) \equiv \delta_D(\br-\br_1)
\eeq
and for $n \geq 2$:
\beqa
\lefteqn{ F_n(\br;\br_1,...,\br_n) \equiv \int \d\br_1' \br_2' \; \tK(\br;\br_1',\br_2')} \nonumber \\ & & \times \sum_{m=1}^{n-1} F_m(\br_1';\br_1,..,\br_m) F_{n-m}(\br_2';\br_{m+1},..,\br_n)
\label{Fn2}
\eeqa
The expression (\ref{Fn1}) clearly shows that the expansion (\ref{exp1}) is a perturbative expansion over powers of the linear mode $\eta$. It is also apparent from eq.(\ref{Fn1}) that $\eta$ should only consist of the linear growing mode. Indeed, the decaying mode would lead to divergences in eq.(\ref{Fn1}) as $\tau \rightarrow -\infty$. This is not surprising since at early times the decaying mode becomes increasingly large so that a perturbative approach fails. Of course, in principle it is possible to include decaying modes if we set up the initial conditions at a finite time $\tau_i$ (i.e. $t_i>0$). Note that in this case we cannot use the kernel $\tK$ derived from eq.(\ref{tK1}). This can be seen from the fact that if we only include the growing mode but set up the initial conditions at $t_i>0$ we should recover the same results for $t>t_i$. However, this cannot be achieved from eq.(\ref{rec2}) using the same kernel $\tK$ since the integration range over $\tau_1$ and $\tau_2$ is changed to $\tau_1>\tau_i$ and $\tau_2>\tau_i$.

Fortunately, for practical purposes the decaying modes should play no role: $t_i \ll t_0$ so that the decaying modes become negligible before one enters the non-linear regime. Hence it is convenient to restrict $\eta(\br)$ to the linear growing mode and to take $t_i=0$.

\subsection{Non-perturbative integral equation}
\label{Integral equation}

Here, we note that the expression (\ref{Fn1}) also shows that the expansion (\ref{exp1}) can be formally reinterpreted as a perturbative expansion over powers of the ``coupling constant'' $g$. Thus, the recursive solution defined by eq.(\ref{exp2}) and eq.(\ref{rec2}) is also the perturbative solution of the integral equation:
\beq
\df(\br) = \eta(\br) + g \int \d\br_1 \d\br_2 \; \tK(\br;\br_1,\br_2) \; \df(\br_1) \df(\br_2)
\label{int1}
\eeq
where the perturbative parameter is $g$. Moreover, if we apply to both sides of eq.(\ref{int1}) the operator $\cO$, using eq.(\ref{tK0}), we obtain eq.(\ref{Bol6}). This means that the solution $\df[\eta]$ of eq.(\ref{int1}) is also a solution of the non-perturbative equation (\ref{Bol6}). Hence eq.(\ref{int1}) is actually equivalent to eq.(\ref{Bol6}) supplemented by the initial condition $\df \rightarrow \eta$ at early times $\tau \rightarrow -\infty$. Indeed, eq.(\ref{Bol6}) admits only one solution (i.e., the distribution function $\df$ is completely specified by its value at some initial time $t_i$ since eq.(\ref{Bol6}) involves a first-order time derivative). Thus, the problem of solving the collisionless Boltzmann equation with this initial condition comes down to look for the solution $\df$ of eq.(\ref{int1}) for a given linear mode $\eta$. We stress here that eq.(\ref{int1}) is {\it non-perturbative}, since it is equivalent to eq.(\ref{Bol6}). Note that in our case we shall eventually put $g=1$. In particular, the relation (\ref{int1}) does not assume that the distribution function $\df$ can be written as a perturbative series over the linear mode $\eta$, that is eq.(\ref{int1}) remains valid even if the perturbative series is only asymptotic (i.e. divergent).

The eq.(\ref{int1}) is similar to the integral form of usual stochastic differential equations (e.g., Langevin equation) where $\eta$ would be called a noise term. Moreover, this formulation is very convenient since it reduces the three equations we started from (Boltzmann eq., Poisson eq., initial conditions) to the only one eq.(\ref{int1}). Thus, the initial conditions $\eta(\br)$ are encoded within the ``evolution equation'' itself, which fully determines the properties of the distribution function $\df$ once the characteristics of $\eta(\br)$ are specified. Here we note that a quadratic equation of the form (\ref{int1}) was derived in \cite{Scoc2} from the usual hydrodynamical description. However, the latter formulation breaks down when shell-crossing occurs while eq.(\ref{int1}) remains valid in the non-linear regime.

\subsection{Diagrams}
\label{Diagrams}

\begin{figure}[htb]

\centerline{\epsfxsize=8 cm \epsfysize=3.5 cm \epsfbox{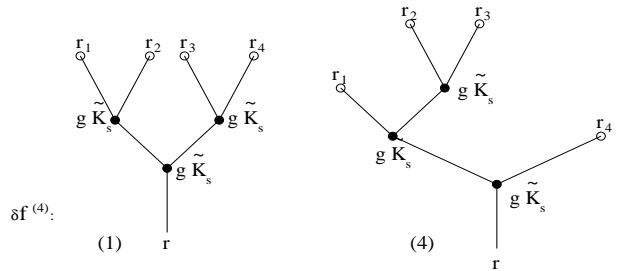}}

\caption{The two diagrams which give the fourth-order term $\df^{(4)}(\br)$. The filled circles correspond to the three-leg vertices, which contribute a weight $g \tKs$ for each one, and the circles represent the external points $\br_1,..,\br_n$. The numbers $(1)$ and $(4)$ give the multiplicity $a_{4,\alpha}$ of each diagram.}
\label{figdiag}

\end{figure}

The perturbative expansion (\ref{rec2}) can be conveniently written as a sum of tree-diagrams with a three-leg vertex. To do so, we first define the symmetrized kernel $\tKs$ by:
\beq
\tKs(\br;\br_1,\br_2) \equiv \frac{1}{2} \; \left[ \tK(\br;\br_1,\br_2) + \tK(\br;\br_2,\br_1) \right] .
\label{tKs}
\eeq
Thus, the relations (\ref{rec2}), (\ref{Fn2}) and (\ref{int1}) remain valid after we replace the kernel $\tK$ by $\tKs$. Then, one can write the term $\dfn$ of order $n$ as the sum over all trees built from $(n-1)$ three-leg vertices with $n$ external points $\br_1,..,\br_n$ over which we integrate. For instance, we show in Fig.\ref{figdiag} the two diagrams which give the fourth-order term $\df^{(4)}$. To construct such diagrams one simply applies a recursive procedure of $(n-1)$ steps which builds the tree upward. To go from step $k$ to step $k+1$ one merely chooses an end-point (i.e. with no links upward) and connects to this point two ``sons'' through a three-leg vertex. At the first step one starts from the ``root'' $\br$ while the final end-points which are left are labeled $\br_1,..,\br_n$ (the order is irrelevant). Then, each vertex contributes a weight $g \tKs$ while to each external point is affected a weight $\eta(\br_i)$. Finally, to obtain $\dfn(\br)$ one simply integrates over the points $\br_1,..,\br_n$ and takes the sum over all diagrams. Note that each diagram of topology $\alpha$ needs to be multiplied by a multiplicity factor $a_{n,\alpha}$.

It is instructive to consider the simple case where $\df$ and $\eta$ are mere numbers, rather than functions. Then we can absorb $\tKs$ into $g$ and the integral equation (\ref{int1}) becomes:
\beq
\df = \eta + g \; \df^2 .
\label{int2}
\eeq
Let us note $c_n$ the sum over all multiplicity factors associated with the diagrams of order $n$, that is: $c_n = \sum_{\alpha} a_{n,\alpha}$ (in particular, $c_4=5$). Then, the perturbative expansion (\ref{exp1}) reads:
\beq
\df = \sum_{n=1}^{\infty} g^{n-1} \; c_n \; \eta^n
\label{exp3}
\eeq
with $c_1 \equiv 1$. However, in this simple case we can directly solve the non-linear eq.(\ref{int2}) which yields (the choice among the two roots of the quadratic eq.(\ref{int2}) is determined by the constraint $\df=\eta$ in the limit $g \rightarrow 0$):
\beq
\df = \frac{1-\sqrt{1 - 4 g \eta}}{2 g} .
\label{df2}
\eeq
Then, expanding the square root which appears in (\ref{df2}) we obtain the coefficients $c_n$:
\beq
c_1=1  \hspace{0.2cm} \mbox{and for} \hspace{0.2cm} n \geq 2 : \;  c_n= \frac{2(2n-3)!}{n! (n-2)!} .
\eeq
Note that these coefficients $c_n$ are simply the Catalan numbers. From the expression (\ref{df2}) we can see that the perturbative expansion (\ref{exp3}) diverges for $|g \eta|>1/4$. Note that this occurs rather early as the second-order term is only one fourth of the first-order term. This means in particular that we cannot use a local approximation for eq.(\ref{int1}), which would consist of taking the mean of eq.(\ref{int1}) over a small phase-space volume $V$ and approximating the integral in the r.h.s. by its value when $\br_1$ and $\br_2$ are restricted to $V$. Of course, this result does not imply that the exact perturbative series diverges for large $\eta$ (i.e. in the non-linear regime), however it suggests that such a behaviour would be quite likely.

\section{Explicit calculations up to the second-order}
\label{Explicit calculations}

We have described in the previous section the general structure of the evolution equation which governs the dynamics of the distribution function $\df(\br)$. In particular, we have shown how we can obtain a perturbative expansion of $\df(\br)$ over powers of the linear growing mode $\eta(\br)$. In this section, we apply these results to the specific case of gravitational clustering in an expanding universe (with $\Om=1$).

\subsection{Linear growing mode}
\label{Linear growing mode}

First, we have to obtain the linear growing mode $\eta(\br)$ (as explained in Sect.\ref{Perturbative expansion} we do not need the decaying modes). This function $\eta(\br)$ must belong to the kernel of the operator $\cO$, see eq.(\ref{ker}), and grow with time. To get $\eta(\br)$ one can simply consider the linear growing mode of the hydrodynamical approximation. Thus, in this approximation all particles at a given point $\bx$ have the same peculiar velocity $\bv$ defined by:
\beq
\bv = a \dot{\bx} , \hspace{0.2cm} \mbox{hence} \hspace{0.2cm} \bp=a \bv .
\label{v1}
\eeq
In the linear regime the growing modes of the density and velocity fields are related by (\cite{Peebles1}):
\beq
\bv_L(\bk,\tau) = i \frac{a}{D_+} \dot{D}_+ \frac{\bk}{k^2} \; \delta_L(\bk,\tau) = i \; \dot{a} \frac{\bk}{k^2} \; \delta_L(\bk,\tau)
\eeq
where the growth factor $D_+(t)$ verifies $D_+(t) \propto a(t)$ since we consider a critical density universe. The dot over ``$\dot{D_+}$'' or ``$\dot{a}$'' stands for the time-derivative ``$\d D_+/\d t$'' or ``$\d a/\d t$''. The subscript $L$ refers to the ``linear'' regime. Thus, using eq.(\ref{v1}) we have in the linear regime:
\beq
\bp_L(\bk,\tau) = i \; \dot{a} a \; \frac{\bk}{k^2} \; \delta_L(\bk,\tau) .
\label{p2}
\eeq
Moreover, since $\delta_L(\bk,\tau) \propto a$ we can write:
\beq
\delta_L(\bk,\tau) = e^{2\tau} \dLo(\bk) , \hspace{0.2cm} \bp_L(\bk,\tau) = i \frac{2}{3} e^{3\tau} \dLo(\bk) \frac{\bk}{k^2} .
\label{p3}
\eeq
Thus the linear growing mode which is solution of the hydrodynamical equations is:
\beq
\df_{\rm hydro}^{(1)}(\bx,\bp,\tau) = (1+\delta_L)(\bx,\tau) \delta_D[ \bp - \bp_L(\bx,\tau)] - \delta_D(\bp) .
\label{hydro}
\eeq
The Dirac distribution $\delta_D[ \bp - \bp_L(\bx,\tau)]$ corresponds to the hydrodynamical description. However, one can check that this function $\df_{\rm hydro}^{(1)}$ is not a solution of eq.(\ref{ker}). To obtain the correct linear mode $\eta(\br)$ which belongs to the kernel of $\cO$ one simply needs to keep only the linear term in eq.(\ref{hydro}). Thus, we write:
\beq
\df^{(1)}(\bx,\bp,\tau) = \delta_L(\bx,\tau) \delta_D(\bp) - \bp_L(\bx,\tau) . \frac{\pl \delta_D}{\pl \bp}(\bp) .
\eeq
In Fourier space we obtain for $\eta(\br) \equiv \df^{(1)}(\br)$:
\beq
\eta(\br) = e^{2\tau} \dLo(\bk) \delta_D(\bp) - i \frac{2}{3} e^{3\tau} \dLo(\bk) \frac{\bk}{k^2} . \frac{\pl \delta_D}{\pl \bp}(\bp) .
\label{eta1}
\eeq
Then, one can easily check that $\cO.\eta=0$ with $\eta$ given by eq.(\ref{eta1}) and $\cO$ by eq.(\ref{O1}).

Here, we can note that up to first-order our system can be described as a pressure-less fluid. Indeed, we have not included any velocity dispersion in the initial conditions. As is well-known, caustics will form as the density field evolves and multi-streaming regions appear. The perturbative approach should break down at this stage since it cannot go beyond these singularities. Thus, we shall check in the following sections that our perturbative treatment recovers the predictions of the standard perturbative approach applied to the equations of motion derived within the hydrodynamic approximation (see \cite{Peebles1} or \cite{Goroff1} for a presentation of this method). This clearly implies that the perturbative method cannot handle the multi-streaming regime. However, we stress that the equations of motion (\ref{Bol1}), (\ref{Bol5}) and (\ref{int1}) remain valid in the non-linear regime after multi-streaming appears.

In order to handle the multi-streaming regime in a perturbative way, one may try to add a small velocity dispersion to the initial conditions. Indeed, this will smooth out the caustics encountered for a pressure-less fluid. Such an approach is investigated in \cite{Dom1} and \cite{Dom2} for instance. However, it is clear that this is not sufficient to go beyond shell-crossing. Indeed, as non-linear structures form, shocks will appear around virialized objects. This can be seen from the analytic solution of gaseous spherical collapse in \cite{Bert1}. Then, a perturbative approach cannot give any information on the inner region beyond the shock. In particular, one cannot rely on a simple equation of state of the form $p=p(\rho)$ for the pressure. Moreover, it is well-known that collisionless and gaseous systems can exhibit qualitatively different behaviours (e.g., existence of Landau damping and phase-mixing for the former, see \cite{Bin1}). This is why it is important to devise methods which are based on the collisionless Boltzmann equation (\ref{Bol1}). This article is a first step in this direction.

\subsection{Second-order term. Comparison with the hydrodynamical approach}
\label{Second-order term}

Once the linear growing mode $\eta(\br)$ is specified we can derive the higher-order terms from the recursion (\ref{rec2}). Thus, using the expression (\ref{eta1}) and eq.(\ref{tK2}) obtained in App.\ref{Calculation of the kernel} for the kernel $\tK$ we get after a simple calculation:
\beqa
\lefteqn{ \df^{(2)}(\bk,\bp,\tau) = \int \d\bk_1 \; \dLo(\bk_1) \dLo(\bk-\bk_1) } \nonumber \\ & & \times \biggl \lbrace \left(1+\frac{2}{3} \frac{\bk.(\bk-\bk_1)}{|\bk-\bk_1|^2} \right) \left[ \frac{3}{7} e^{4 \tau} \frac{\bk.\bk_1}{k_1^2} \delta_D(\bp) \right. \nonumber \\ & & \hspace{0.7cm} \left. - \frac{2i}{5} e^{5 \tau} \frac{\bk_1}{k_1^2} . \frac{\pl \delta_D}{\pl \bp}(\bp) - \frac{6i}{35} e^{5 \tau} \frac{\bk.\bk_1}{k_1^2} \frac{\bk}{k^2} . \frac{\pl \delta_D}{\pl \bp}(\bp) \right] \nonumber \\ & & \hspace{0.5cm} - \frac{2}{9} e^{6 \tau} \frac{\bk_1}{k_1^2} . \frac{\pl}{\pl \bp} \left( \frac{\bk-\bk_1}{|\bk-\bk_1|^2} . \frac{\pl \delta_D}{\pl \bp}(\bp) \right) \biggl \rbrace .
\label{d2f1}
\eeqa
To derive eq.(\ref{d2f1}) we used identities of the form:
\beq
F(\bk.\bp) \; \delta_D(\bp) = F(0) \; \delta_D(\bp)
\label{Dirac1}
\eeq
and:
\beqa
F(\bk_1.\bp) \frac{\bk}{k^2} . \frac{\pl \delta_D}{\pl \bp}(\bp) & = & F(0) \frac{\bk}{k^2} . \frac{\pl \delta_D}{\pl \bp}(\bp) \nonumber \\ & & - F'(0) \frac{\bk.\bk_1}{k^2} \; \delta_D(\bp)
\label{Dirac2}
\eeqa
which are valid for any function $F$. This allows us to eliminate the functions $\psi$ which appear in $\tK$ in eq.(\ref{tK2}) since we can express all factors in terms of $\psi(s,0)=1/s$, see eq.(\ref{psi3}).

It is interesting to compare this result for $\df^{(2)}$ with the results obtained within the usual hydrodynamical approach. In this latter case, the system is fully defined by the density contrast and the mean fluid velocity at each point $\bx$ (there is no velocity dispersion). Thus, we need to derive the density and the mean collective velocity from eq.(\ref{d2f1}).

\subsubsection{Density contrast}
\label{Density contrast}

From eq.(\ref{delta1}) and eq.(\ref{d2f1}) we obtain the second-order term for the density contrast:
\beqa
\lefteqn{ \delta^{(2)}(\bk,\tau) = e^{4 \tau} \int \d\bk_1 \; \dLo(\bk_1) \dLo(\bk-\bk_1) } \nonumber \\ & & \times \left( \frac{3}{7} \frac{\bk.\bk_1}{k_1^2} + \frac{2}{7} \frac{\bk.\bk_1}{k_1^2} \frac{\bk.(\bk-\bk_1)}{|\bk-\bk_1|^2} \right) .
\eeqa
Going back to real space, we get after a change of variables:
\beqa
\lefteqn{ \delta^{(2)}(\bx,\tau) = e^{4 \tau} \int \d\bk_1 \d\bk_2 \; e^{i(\bk_1+\bk_2).\bx} \; \dLo(\bk_1) \dLo(\bk_2) } \nonumber \\ & & \times \left[ \frac{5}{7} + \frac{\bk_1.\bk_2}{k_1^2} + \frac{2}{7} \left( \frac{\bk_1.\bk_2}{k_1 k_2} \right)^2 \right] .
\eeqa
Thus, we recover the result obtained by the hydrodynamical approach (\cite{Peebles1}).

We can note that this result agrees with the calculations performed in \cite{Bha1} from the BBGKY hierarchy. Indeed, in that paper it was shown that the lowest order non-linear correction to the two-point correlation function derived from the BBGKY hierarchy (which remains valid in the non-linear regime after shell-crossing) matches the prediction of the standard hydrodynamical approach. Note that the BBGKY hierarchy can be derived from the collisionless Boltzmann equation (\ref{Bol1}), see \cite{Peebles1}. Therefore, both works explicitly show that multi-streaming effects cannot be handled through a perturbative method.

\subsubsection{Peculiar velocity}
\label{Peculiar velocity}

From the distribution function $\df$ we can also derive the mean velocity of the fluid. Indeed, the hydrodynamics equations are obtained from moments of the Boltzmann equation. In particular, the hydrodynamic peculiar velocity is defined by:
\beq
\rho \bv \equiv \int \d \bp \; f(\bx,\bp,\tau) \; \frac{\bp}{a} .
\label{vh1}
\eeq
Since $\rho=\rhob(1+\delta)$ we obtain using eq.(\ref{df1}):
\beq
\left[ 1 + \delta(\bx,\tau) \right] \bv(\bx,\tau) = e^{-2\tau} \int \d \bp \; \df(\bx,\bp,\tau) \; \bp
\label{vh2}
\eeq
which yields for the second-order term:
\beq
\bv^{(2)} = e^{-2\tau} \int \d \bp \; \df^{(2)} \; \bp \; - \; \delta^{(1)} \bv^{(1)} .
\label{vh3}
\eeq
The first-order term is simply $\bv^{(1)}=\bv_L=e^{-2\tau}\bp_L$ which is given by eq.(\ref{p3}). Then, we obtain the second-order term from eq.(\ref{d2f1}). This yields:
\beqa
\lefteqn{ \bv^{(2)}(\bk,\tau) = i \; e^{3 \tau} \int \d\bk_1 \; \dLo(\bk_1) \dLo(\bk-\bk_1) } \nonumber \\ & & \times \left[ \frac{-4}{15} \frac{\bk_1}{k_1^2} + \frac{4}{15} \frac{\bk.(\bk-\bk_1)}{|\bk-\bk_1|^2} \frac{\bk_1}{k_1^2} + \frac{6}{35} \frac{\bk.\bk_1}{k_1^2} \frac{\bk}{k^2} \right. \nonumber \\ & & \hspace{0.6cm} \left. + \frac{4}{35} \frac{\bk.\bk_1}{k_1^2} \frac{\bk.(\bk-\bk_1)}{|\bk-\bk_1|^2} \frac{\bk}{k^2} \right] .
\label{vh4}
\eeqa
Within the hydrodynamical approach, when one expands the density and velocity fields over the linear growing mode (i.e. there are no decaying modes) the velocity field is nonrotational (e.g., \cite{Peebles1}). Then it is fully defined by its divergence and one usually introduces:
\beq
\theta(\bx,\tau) \equiv \frac{\nabla . \bv}{\dot a} , \hspace{0.2cm} \mbox{hence}\hspace{0.2cm} \theta(\bk,\tau) = \frac{3}{2} \; e^{\tau} \; i \bk.\bv(\bk,\tau) .
\label{theta1}
\eeq
In real space, using eq.(\ref{vh4}) we recover the result of the hydrodynamical approach (\cite{Peebles1}):
\beqa
\lefteqn{ \theta^{(2)}(\bx,\tau) = - e^{4 \tau} \int \d\bk_1 \d\bk_2 \; e^{i(\bk_1+\bk_2).\bx} \; \dLo(\bk_1) \dLo(\bk_2) } \nonumber \\ & & \times \left[ \frac{3}{7} + \frac{\bk_1.\bk_2}{k_1^2} + \frac{4}{7} \left( \frac{\bk_1.\bk_2}{k_1 k_2} \right)^2 \right] .
\label{theta2}
\eeqa
In a similar fashion, we can consider the normalized vorticity $\bom \equiv (\nabla \times \bv)/{\dot a}$. Using eq.(\ref{vh4}) we can easily check that we obtain $\bom^{(2)}=0$ (and of course the linear mode as defined in eq.(\ref{p3}) is nonrotational). Thus, we see that up to the second-order we recover the results of the hydrodynamical approach.

\section{Divergence of perturbative series}
\label{Divergence of perturbative series}

In the previous section we noticed that the perturbative treatment of the collisionless Boltzmann equation yields the same results at second-order as the hydrodynamical approach when the linear mode is given by eq.(\ref{eta1}). In fact, it is easy to see that this equality extends to all orders of perturbation theory. Indeed, let us consider an initial condition of the form (\ref{eta1}) which is very smooth. That is, $\dLo(\bk)$ shows no power at small scales (i.e. large wavenumbers $k$) and the flow is nonrotational, defined by the linear growing mode of hydrodynamics. Then, there is no shell-crossing until a finite time $\tau_c$ (i.e. $t_c>0$). This means that the hydrodynamical description is actually exact at early times $t<t_c$: at each point there is only one velocity. Then, until the non-zero time $t_c$ the solution of hydrodynamics (i.e. continuity and Euler equations) is also a solution of the collisionless Boltzmann equation from which they derive. This implies that the density contrast and velocity fields obtained by the perturbative theory in both frameworks are identical. Moreover, the form of the perturbative expansion does not depend on the properties of $\dLo(\bk)$, hence this identity holds for any initial conditions $\dLo(\bk)$. As a consequence, if the linear mode is of the form of eq.(\ref{eta1}) the perturbative results obtained from the collisionless Boltzmann equation and from the hydrodynamical description are identical. Here, we note that this identity was explicitly derived at all orders within the framework of the Zel'dovich approximation in \cite{Bha2}. Therefore, in order to study the non-linear regime one needs to devise non-perturbative methods applied to the collisionless Boltzmann equation (\ref{Bol1}).

On the other hand, in the case of CDM-like power-spectra (or power-law power-spectra) there is some power at all scales. In particular, the density fluctuations increase with no limit at small-scales (if we do not include a cutoff at small scales). Then, as soon as $t>0$ there is some shell-crossing. This implies that the hydrodynamical solution is no longer a solution of the collisionless Boltzmann equation (actually, the former is not well defined). Then, the perturbative expansion must diverge (otherwise the hydrodynamical solution and the Boltzmann solution would be identical since they would be equal to the same perturbative solution). Hence, we conclude that {\it for hierarchical initial conditions the perturbative expansion is only asymptotic}. If we include a cutoff at small scales for the initial power-spectrum the perturbative series may converge until the finite time $t_c$ when shell-crossing occurs for the first time (though this is not guaranteed a priori). However, it must diverge at later times which implies that the perturbative series diverges at all times of practical interest.

In fact, the system we investigate here can actually be even more pathological than these results suggest. First, as we show in a companion paper (\cite{paper2}) where we develop a non-perturbative method to study the quasi-linear regime, the generating function of the cumulants of the density contrast obtained at leading order has a radius of convergence which is zero independently of shell-crossing for $P(k) \propto k^n$ with $n<0$. This means that the high-density tail of the probability distribution function of the density contrast cannot be evaluated by perturbative means in this case. Secondly (and more importantly), the terms encountered in perturbative expansions actually diverge beyond some finite order for power-law power-spectra (e.g., \cite{Scoc1}, \cite{paper5}). This implies that one can only use perturbation theory up to a finite order at best.

\section{Other cosmologies}
\label{Other cosmologies}

In this section, we show how we can extend the perturbative calculations developed in Sect.\ref{Building a perturbative expansion} for a critical density universe to arbitrary cosmological parameters. In the general case, the scale factor $a(t)$ is no longer a power-law hence it is not useful to introduce the time coordinate $\tau$ as in eq.(\ref{tau}). Thus, from eq.(\ref{Bol4}) and keeping the coordinate $t$ we define the operator $\cO$ by:
\beqa
\lefteqn{ (\cO . \df) (\bk,\bp,t) \equiv \frac{\pl \df}{\pl t} + i \frac{\bk.\bp}{a^2} \; \df } \nonumber \\ & & + i \frac{4\pi\cG\rhob}{a} \frac{\bk}{k^2} . \frac{\pl \delta_D}{\pl \bp}(\bp) \int \d\bp' \; \df(\bk,\bp',t) .
\label{O2}
\eeqa
Next, as in Sect.\ref{Linear growing mode} we obtain the growing mode $\eta(\bk,\bp,t)$ by keeping only the linear terms of the linear hydrodynamical growing mode which can still be written as in eq.(\ref{hydro}). This yields:
\beq
\eta = D_+ \dLo(\bk) \delta_D(\bp) - i \; a^2 \dot{D}_+ \dLo(\bk) \frac{\bk}{k^2} . \frac{\pl \delta_D}{\pl \bp}(\bp)
\label{eta2}
\eeq
where $D_+(t)$ is the usual linear growing mode, which is no longer proportional to $a(t)$. Then, one can easily check that $\cO.\eta=0$ with $\eta$ given by eq.(\ref{eta2}) and $\cO$ by eq.(\ref{O2}), using the fact that $D_+(t)$ obeys (\cite{Peebles1}):
\beq
\ddot{D}_+ + 2 \frac{\dot a}{a} \dot{D}_+ = \frac{4\pi\cG\rhob}{a^3} \; D_+ .
\eeq
Then, in order to get the second-order term we simply keep the structure of the result (\ref{d2f1}) obtained for the critical density universe but we replace all time-dependent factors by unknown functions. Thus, we write:
\beqa
\lefteqn{ \df^{(2)}(\bk,\bp,t) = \int \d\bk_1 \; \dLo(\bk_1) \dLo(\bk-\bk_1) } \nonumber \\ & & \times \biggl \lbrace g_1 \frac{\bk.\bk_1}{k_1^2} \delta_D(\bp) + g_2 \frac{\bk.\bk_1}{k_1^2} \frac{\bk.(\bk-\bk_1)}{|\bk-\bk_1|^2} \delta_D(\bp)  \nonumber \\ & & - i g_3 \frac{\bk_1}{k_1^2} . \frac{\pl \delta_D}{\pl \bp}(\bp) - i g_4 \frac{\bk.(\bk-\bk_1)}{|\bk-\bk_1|^2} \frac{\bk_1}{k_1^2} . \frac{\pl \delta_D}{\pl \bp}(\bp) \nonumber \\ & & - i g_5 \frac{\bk.\bk_1}{k_1^2} \frac{\bk}{k^2} . \frac{\pl \delta_D}{\pl \bp}(\bp) - i g_6 \frac{\bk.\bk_1}{k_1^2} \frac{\bk.(\bk-\bk_1)}{|\bk-\bk_1|^2} \frac{\bk}{k^2} . \frac{\pl \delta_D}{\pl \bp}(\bp) \nonumber \\ & & - g_7 \frac{\bk_1}{k_1^2} . \frac{\pl}{\pl \bp} \left( \frac{\bk-\bk_1}{|\bk-\bk_1|^2} . \frac{\pl \delta_D}{\pl \bp}(\bp) \right) \biggl \rbrace
\label{d2f2}
\eeqa
where $g_1(t),...,g_7(t)$ are functions of time. Next, we substitute eq.(\ref{d2f2}) into eq.(\ref{Bol4}), using eq.(\ref{eta2}) for the first-order term of $\df$, and keeping only the second-order terms we obtain a system of differential equations for the unknown functions $g_1(t),...,g_7(t)$. This yields:
\beqa
& & a^2 \dg_1 = g_3+g_5 \; , \;\; a^2 \dg_2 = g_4+g_6 \; , \;\; a \dg_3 = 4\pi\cG\rhob D_+^2 \\ & & a^2 \dg_4 = 2 g_7 \; , \;\; a \dg_5 = 4\pi\cG\rhob g_1 \\ & & a \dg_6 = 4\pi\cG\rhob g_2 \; , \;\; \dg_7 = 4\pi\cG\rhob D_+^2
\eeqa
which determines the functions $g_1(t),...,g_7(t)$. The boundary condition at $t=0$ is given by the constraint that for $t \rightarrow 0$ we must recover the result (\ref{d2f1}) obtained for the critical density universe. We shall not go further here since the discussion of Sect.\ref{Divergence of perturbative series} remains valid for arbitrary cosmologies hence we already know that we shall eventually recover the results obtained through the hydrodynamical approach, described in \cite{Bouchet1}.

Thus, in order to derive the perturbative expansion of the distribution function $\df$ in the general case one first computes the terms $\df^{(n)}$ up to the required order for a critical density universe, using the method developed in Sect.\ref{Building a perturbative expansion}. This provides an explicit expression for $\df^{(n)}$. Then, one looks for a solution of eq.(\ref{Bol4}) in the case of interest (which differs from the critical density universe) by keeping the same structure for $\df^{(n)}$ but replacing all time-dependent factors by unknown functions of time. Then, these functions are determined by substitution into eq.(\ref{Bol4}). The need for this two-step procedure comes from the fact that in the general case the calculation of the operator $\cOm$ is not straightforward. More precisely, in order to derive $\cOm$ one can still follow the procedure outlined in App.\ref{Calculation of the exponential} up to eq.(\ref{Kn2}). However, the recursion obeyed by the functions $R_n(\sigma)$ is no longer a simple Laplace convolution so that it is not obvious to find a simple analytic solution.

\section{Functional formulation}
\label{Functional formulation}

Finally, we briefly describe how one can use the integral equation (\ref{int1}) to obtain an explicit expression for the statistical properties of the distribution function. In this section, we shall work in real space $\bx$, and we note $\bom$ the 7-dimensional coordinate $\bom=(\bx,\bp,\tau)$. Thus, the fields $\eta(\bom)$ and $\df(\bom)$ are real. Taking the Fourier transform of eq.(\ref{int1}) we see that $\df(\bom)$ and $\eta(\bom)$ are related by the same quadratic equation where the kernel $\tK(\br;\br_1,\br_2)$ is replaced by its real-space Fourier transform $\tK(\bom;\bom_1,\bom_2)$. In practice we are not really interested in the exact solution $\df$ associated with a given field $\eta$. Indeed, since the initial condition $\eta$ is a random field we are only interested in the statistical properties of the distribution function $\df$. These can be derived from the functional $Z[j]$ of the test field $j(\bom)$ defined by:
\beq
Z[j] \equiv \lag e^{\int \d\bom \; j.\df} \rag
\label{Z1}
\eeq
where $\lag .. \rag$ expresses the average over the initial conditions. Expanding the exponential we get:
\beqa
Z[j] & = & 1 + \sum_{n=1}^{\infty} \frac{1}{n!} \int \d\bom_1 .. \d\bom_n \; j(\bom_1) .. j(\bom_n) \nonumber \\ & & \times \; \lag \df(\bom_1) .. \df(\bom_n) \rag
\label{Z2}
\eeqa
which clearly shows that $Z[j]$ yields all moments $\lag \df(\bom_1) .. \df(\bom_n) \rag$ of the distribution function $\df$. If we assume the initial conditions to be Gaussian we can write the average (\ref{Z1}) as the path-integral:
\beq
Z[j] = \cNo \int [\d\eta(\bom)] \; e^{j.\df - \frac{1}{2} \eta . \Dem . \eta}
\label{Z3}
\eeq
where we introduced the normalization constant:
\beq
\cNo \equiv \left( \Det \Dem \right)^{1/2} .
\eeq
Here $\df[\eta]$ must be understood as the solution of eq.(\ref{int1}) (in real $\bom$-space) associated with a given $\eta$. We also used the short-hand notations: $j.\df \equiv \int \d\bom \; j(\bom) . \df(\bom)$ and $\eta . \Dem . \eta \equiv \int \d\bom_1 \d\bom_2 \; \eta(\bom_1) . \Dem(\bom_1,\bom_2) . \eta(\bom_2)$. The kernel $\Dem$ is the inverse of the kernel $\De$ defined by:
\beq
\De(\bom_1,\bom_2) \equiv \lag \eta(\bom_1) \eta(\bom_2) \rag
\label{De1}
\eeq
which fully determines the statistics of the random field $\eta(\bom)$ since the latter is assumed to be Gaussian. Here we must note that the kernel $\De$ is not invertible. This is related to the fact that the linear mode $\eta(\bom)$ is restricted to the form (\ref{eta1}), which also ensures that $\cO . \eta = 0$. Thus, in eq.(\ref{Z3}) it is understood that the kernels $\De$ and $\Dem$ are regularized through a small modification described by an infinitesimal parameter $\epsilon$, which makes $\De$ and $\Dem$ invertible. For instance, one can add to the kernel $\De$ a term of the form $\epsilon \times \delta_D(\bom_1-\bom_2)$. Then, the final results are obtained by taking the limit $\epsilon \rightarrow 0$ in the calculations. Using eq.(\ref{int1}) we can make the change of variable $\eta \rightarrow \df$ in the expression (\ref{Z3}). This yields:
\beqa
Z[j] & = & \cNo \int [\d\df(\bom)] \; | \Det M | \nonumber \\ & & \times \; e^{j.\df - \frac{1}{2} (\df- g \tK \df^2) . \Dem . (\df - g \tK \df^2)}
\label{Z4}
\eeqa
where we used the notation:
\beq
(\tK \df^2)(\bom) \equiv \int \d\bom_1 \d\bom_2 \tK(\bom;\bom_1,\bom_2) \df(\bom_1) \df(\bom_2) .
\eeq
In eq.(\ref{Z4}) we introduced the matrix $M$ defined by:
\beqa
\lefteqn{ M(\bom_1,\bom_2) \equiv \frac{\delta \eta(\bom_1)}{\delta ( \df(\bom_2) )} } \nonumber \\ & & = \delta_D(\bom_1 - \bom_2) - 2 g \int \d \bom' \; \tK_s(\bom_1;\bom_2,\bom') \df(\bom') .
\label{M1}
\eeqa
Here we used eq.(\ref{int1}) (in real $\bom$-space) and $\tK_s$ is the symmetrized part of $\tK$, as in eq.(\ref{tKs}). Next, we need to compute the Jacobian $|\Det M|$. To do so, we use the relation:
\beq
\Det M = e^{{\rm Tr} \; \ln M}
\eeq
and from eq.(\ref{M1}) we have:
\beq
\Tr \; \ln M = - \sum_{n=1}^{\infty} \frac{(2 g)^n}{n} \; \Tr \; A^n
\eeq
where we introduced the matrix $A$:
\beq
A(\bom_1,\bom_2) \equiv \int \d \bom' \; \tK_s(\bom_1;\bom_2,\bom') \; \df(\bom') .
\eeq
Hence a simple recursion yields for $n \geq 2$:
\beqa
\lefteqn{ A^n(\bom,\bom') = \int \d\bom_1 ..\d\bom_{n-1} \d\bom_1' ..\d\bom_{n}' \; \df(\bom_1') .. \df(\bom_n') } \nonumber \\ & & \times \; \tK_s(\bom;\bom_1,\bom_1') \tK_s(\bom_1;\bom_2,\bom_2') .. \tK_s(\bom_{n-1};\bom',\bom_n') .
\label{A1}
\eeqa
Next, we note from eq.(\ref{tK2}) that the kernel $\tK_s(\bom;\bom_1,\bom_2)$ involves a factor $\theta(\tau-\tau_1) \delta_D(\tau_2-\tau_1)$ (which also ensures that causality is satisfied). Then, we see that the integrand in eq.(\ref{A1}) involves a factor $\theta(\tau-\tau_1) .. \theta(\tau_{n-1}-\tau')$. Thus, the Heaviside factors imply that the time-coordinates must obey: $\tau>\tau_1>..>\tau_{n-1}>\tau'$ in order that the integrand be non-zero. This also implies that $A^n(\bom,\bom)=0$ for $n \geq 2$. On the other hand, the trace $\Tr A^n$ is defined by:
\beq
\Tr \; A^n \equiv \int \d\bom \; A^n(\bom,\bom) .
\label{T1}
\eeq
Hence we see that $\Tr A^n=0$ for $n\geq 2$. Note that this result only relies on causality requirements (which translate into the Heaviside factors) and it also applies to usual stochastic differential equations (e.g., Langevin equation, see \cite{Zinn1}). Thus, we are left with $\Det M = \exp(-2 g \Tr A)$, with:
\beqa
\lefteqn{ 2 \Tr A = 2 \int \d\bom \d\bom' \; \tK_s(\bom;\bom,\bom') \; \df(\bom') } \nonumber \\ & & = \int \d\bom \d\bom' \; \left[ \tK(\bom;\bom,\bom') + \tK(\bom;\bom',\bom) \right] \; \df(\bom') .
\eeqa
Then, from eq.(\ref{tK2}) we can see that:
\beq
\int \d\bom \; \tK(\bom;\bom,\bom') = \int \d\bom \; \tK(\bom;\bom',\bom) = 0 .
\label{T2}
\eeq
Indeed, the second term in eq.(\ref{tK2}) vanishes since $\tau=\tau_1$ and we can push the integration path over $s_1$ to the right: Re$(s_1) \rightarrow \infty$. The first term vanishes after we integrate over $s_1$ and next over $\bp$ or over the angles of $\bk'$ (using the fact that $\tK(\bom;\bom',\bom) \propto \delta_D(\bk')$). Note that this latter integration involves the ``regularization'' of the gravitational interaction discussed in Sect.\ref{Equations of motion} below eq.(\ref{Bol1}). Indeed, as discussed in Sect.\ref{Equations of motion} apparently undetermined integrals are given a meaning by the prescription that one must first perform the integration over angles (or introduce a large-scale cutoff $k_c \rightarrow 0$). Thus, we get $\Det M=1$ and we can write eq.(\ref{Z4}) as:
\beqa
Z[j] = \cNo \int [\d\df(\bom)] e^{j.\df - \frac{1}{2} \df . \Dem . \df + \frac{g}{3!} K_3 \df^3 - \frac{g^2}{4!} K_4 \df^4} \nonumber \\ 
\label{Z5}
\eeqa
Here we used the short-hand notations:
\beqa
K_3 \; \df^3 & \equiv & \sum_{\mbox{perm.}} \int \d\bom \d\bom_1 .. \d\bom_3 \; \df(\bom_1) \df(\bom_2) \df(\bom_3) \nonumber \\ & & \times \; \Dem(\bom_1,\bom) \tK(\bom;\bom_2,\bom_3)
\label{K3}
\eeqa
and:
\beqa
\lefteqn{ K_4 \df^4 \equiv \frac{1}{2} \sum_{\mbox{perm.}} \int \d\bom \d\bom' \d\bom_1 .. \d\bom_4 \; \df(\bom_1) \df(\bom_2) \df(\bom_3) } \nonumber \\ & & \times \; \df(\bom_4) \tK(\bom;\bom_1,\bom_2) \Dem(\bom,\bom') \tK(\bom';\bom_3,\bom_4) 
\label{K4}
\eeqa
where the sums run over the permutations of $(\bom_1,\bom_2,\bom_3)$ and $(\bom_1,\bom_2,\bom_3,\bom_4)$. We symmetrized the kernels $K_3$ and $K_4$ in eq.(\ref{K3}) and eq.(\ref{K4}) because it simplifies perturbative expansions. Thus, the path-integral (\ref{Z5}) yields an explicit expression for the functional $Z[j]$, hence for all statistical properties of the distribution function $\df(\bom)$. Indeed, all terms in this expression are known. The kernel $\tK$ which yields $K_3$ and $K_4$ is given in eq.(\ref{tK2}) while the kernel $\Dem$ is obtained from $\De$ (after a suitable regularization). The latter is given by eq.(\ref{De1}) and eq.(\ref{eta1}) and it can be expressed in terms of the power-spectrum $P(k)$ of the linear density field $\dLo(\bk)$.

The correlation functions can be obtained from eq.(\ref{Z5}) as a perturbative series over the coupling constant $g$. Indeed, we noticed in Sect.\ref{Integral equation} that this is equivalent to the usual expansion over the linear growing mode $\eta(\bom)$. To do so we simply develop the exponential term in eq.(\ref{Z5}) over $g K_3$ and $g^2 K_4$ as a series over $g$. This corresponds exactly to the Feynman diagrams in standard Quantum Field Theories for a ``$\phi^3+\phi^4$'' theory (e.g., \cite{Zinn1}) (note that we defined the kernels $K_3$ and $K_4$ in such a way that they are symmetric). One must merely pay attention to the fact that the kernels $K_3$ and $K_4$ are not pure numbers. Of course, this is due to the long-range character of gravity which leads to a non-local theory (while usual Quantum Field Theories are local). Note that we do not need the explicit expression of the inverse kernel $\Dem$. For instance, in order to compute the skewness some terms which depend on $\Dem$ appear but they actually cancel one another.

In practice, the expression (\ref{Z5}) is not very convenient for perturbative calculations and it is easier to use the direct method described in Sect.\ref{Building a perturbative expansion}, or simply use the standard hydrodynamical approach. However, we stress that eq.(\ref{Z5}) presents the advantage to be an explicit non-perturbative expression for the functional $Z[j]$. Thus, one may hope to be able to extract some useful information from this relation for the non-linear regime. However, this is beyond the scope of this article.

\section{Conclusion}
\label{Conclusion}

Thus, in this article we have developed a systematic method to obtain the solution of the collisionless Boltzmann equation, which describes large-scale structure formation in the universe, as a perturbative series. Moreover, we have explained that these perturbative results are identical to those obtained from the hydrodynamical description of the fluid, if we start with the same initial conditions. Thus, multi-streaming cannot be handled through perturbative methods. Besides, for hierarchical scenarios the perturbative expansions diverge. The interest of this approach is that, contrary to the hydrodynamical model, the collisionless Boltzmann equation provides a rigorous description of the dynamics, even in the non-linear regime. Hence in order to tackle the non-linear regime where shell-crossing plays a key role one should use these equations of motions. Then, the results presented in this paper may serve as a basis for a study of the non-linear regime. In particular, we have obtained a non-perturbative integral equation (\ref{int1}) which explicitly relates the non-linear distribution function to the initial conditions. Besides, we have derived an explicit path-integral expression (\ref{Z5}) which yields the correlation functions of the actual non-linear density field at all orders. Although this result provides a formal description of the statistics of the density field one still needs to check whether it is possible to extract some practical information from this expression. Thus, we shall describe in companion papers an alternative non-perturbative method, based on the path-integral (\ref{Z3}), which allows one to derive rigorous results in the quasi-linear regime for the probability distribution of the density contrast for Gaussian initial conditions (\cite{paper2}) as well as for some non-Gaussian scenarios (\cite{paper3}). Such a formalism can also be used in the non-linear regime (\cite{paper4}).

\appendix

\section{Calculation of the exponential $e^{-\sigma \cO}$}
\label{Calculation of the exponential}

In order to obtain the kernel $\tK$ defined in eq.(\ref{tK1}) we first need to compute the exponential $e^{-\sigma \cO}$. Thus, for a given function $F(\br)=F(\bk,\bp,\tau)$ we look for the function $B(\br,\sigma)$ defined by:
\beq
B(\br,\sigma) \equiv e^{-\sigma \cO(\br)} . F
\label{B1}
\eeq
To derive $B(\br,\sigma)$ we note that this function satisfies the initial-value problem:
\beq
\frac{\pl B}{\pl \sigma} = - \cO . B  \hspace{0.3cm} \mbox{and} \hspace{0.3cm} B(\br,0) = F(\br) .
\label{B2}
\eeq
Moreover, we are only interested in the restriction of $B(\br,\sigma)$ to $\sigma \geq 0$. Using eq.(\ref{O1}) we obtain the first-order integro-differential equation:
\beqa
\lefteqn{ \frac{\pl B}{\pl \sigma} + \frac{\pl B}{\pl \tau} = - 3 i (\bk.\bp) \; e^{-\tau} \; B } \nonumber \\  & & - 2 i \; e^{\tau} \; \frac{\bk}{k^2} . \frac{\pl \delta_D}{\pl \bp}(\bp) \int \d\bp' \; B(\bk,\bp',\tau,\sigma) .
\label{B3}
\eeqa
This equation is formally solved by the method of characteristics which yields the integral form:
\beqa
\lefteqn{ B(\bk,\bp,\tau+\sigma,\sigma) = B(\bk,\bp,\tau,0) e^{3i (\bk.\bp) e^{-\tau} [ e^{-\sigma}-1 ] } } \nonumber \\ & & - 2 i \frac{\bk}{k^2} . \frac{\pl \delta_D}{\pl \bp}(\bp) \int_0^{\sigma} \d \sigma' e^{3i (\bk.\bp) e^{-\tau} [ e^{-\sigma}-e^{-\sigma'} ]} \nonumber  \\ & & \hspace{0.4cm} \times \; e^{\tau+\sigma'} \int \d\bp' \; B(\bk,\bp',\tau+\sigma',\sigma') .
\label{B4}
\eeqa
The form of eq.(\ref{B4}) shows that we can consider in the next intermediate steps $\bk$ and $\tau$ to be fixed constants and we are led to introduce the functions:
\beq
G(\bp,\sigma) \equiv B(\bk,\bp,\tau+\sigma,\sigma)
\label{G0}
\eeq
and:
\beq
U(\bp,\sigma) \equiv B(\bk,\bp,\tau,0) \; e^{3i (\bk.\bp) e^{-\tau} [ e^{-\sigma}-1 ] } ,
\eeq
as well as the kernel:
\beqa
\cK(\bp,\sigma|\bp',\sigma') & \equiv & \theta(\sigma-\sigma') \frac{\bk}{k^2} . \frac{\pl \delta_D}{\pl \bp}(\bp) e^{\sigma'} \nonumber \\ & & \times \; e^{3i (\bk.\bp) e^{-\tau} [ e^{-\sigma}-e^{-\sigma'} ]}
\eeqa
where $\theta(\sigma-\sigma')$ is Heaviside's function. Then, eq.(\ref{B4}) reads:
\beq
G(\bp,\sigma) = U(\bp,\sigma) + \nu \int \cK(\bp,\sigma|\bp',\sigma') G(\bp',\sigma') \d\bp' \d\sigma'
\label{G1}
\eeq
where we introduced the constant:
\beq
\nu \equiv - 2 i \; e^{\tau}
\eeq
and we integrate over $\sigma'$ from $0$ to $+\infty$: $\int \d\sigma' \equiv \int_0^{\infty} \d\sigma'$. The linear integral equation (\ref{G1}) is a standard Fredholm equation of the second kind and it can be solved as a perturbative series over $\nu$:
\beqa
\lefteqn{ G(\bp,\sigma) = U(\bp,\sigma) } \nonumber \\ & & + \sum_{n=1}^{\infty} \nu^n \int \cK_n(\bp,\sigma|\bp',\sigma') U(\bp',\sigma') \d\bp' \d\sigma'
\label{G2}
\eeqa
where the kernels $\cK_n$ are defined by the recursion:
\beq
\cK_n \equiv \int \d\bp'' \d\sigma'' \cK(\bp,\sigma|\bp'',\sigma'') \cK_{n-1}(\bp'',\sigma''|\bp',\sigma')
\label{Kn1}
\eeq
with $\cK_1(\bp,\sigma|\bp',\sigma') \equiv \cK(\bp,\sigma|\bp',\sigma')$. Thus, in order to obtain $G(\bp,\sigma)$ we simply need to sum up the series in the r.h.s. of eq.(\ref{G2}). One can check by substitution into eq.(\ref{Kn1}) that these kernels $\cK_n$ are given for $n \geq 2$ by:
\beqa
\cK_n & = & \frac{\bk}{k^2} . \frac{\pl \delta_D}{\pl \bp}(\bp) \left( -3i e^{-\tau} \right)^{n-1} \int_0^{\infty} \d\sigma_{n-1} \; \theta(\sigma-\sigma_{n-1}) \nonumber \\ & & \times \; e^{3i (\bk.\bp) e^{-\tau} [ e^{-\sigma}-e^{-\sigma_{n-1}} ]} \; R_n(\sigma_{n-1})
\label{Kn2}
\eeqa
where we introduced:
\beqa
\lefteqn{R_n(\sigma_{n-1}) \equiv \int_0^{\infty} \d\sigma_1 .. \d\sigma_{n-2} \; \theta(\sigma_{n-1}-\sigma_{n-2}) .. \theta(\sigma_1-\sigma')} \nonumber \\ & & \times \; e^{\sigma'+\sigma_1+..+\sigma_{n-1}} \left[ e^{-\sigma_{n-1}}-e^{\sigma_{n-2}} \right] .. \left[ e^{-\sigma_1}-e^{-\sigma'} \right] \nonumber \\
\label{Rn1}
\eeqa
for $n \geq 3$ and $R_2(\sigma_1) \equiv \theta(\sigma_1-\sigma') \; \left[ e^{\sigma'}-e^{\sigma_1} \right]$. One can see from eq.(\ref{Rn1}) that the functions $R_n(\sigma)$ actually satisfy the recursion:
\beq
R_n(\sigma) =  \int_0^{\sigma} \d l \; \left( 1 - e^{\sigma-l} \right) \; R_{n-1}(l)
\label{Rn2}
\eeq
which has the form of a simple Laplace convolution. This recursion is easily solved by taking the Laplace transform of eq.(\ref{Rn2}) which yields, after going back to $\sigma$-space through an inverse Laplace transform:
\beq
R_n(\sigma) = e^{\sigma'} \inta \frac{\d s}{2\pi i} \; e^{s(\sigma-\sigma')} \; \left[ \frac{-1}{s (s-1)} \right]^{n-1} .
\label{Rn3}
\eeq
Here and in the following calculations the integration path over the Laplace variables like $s$ will always be to the right of the singularities of the integrand. Thus, in eq.(\ref{Rn3}) the integration path for $s$ obeys: $\Re(s)>1$. Then, eq.(\ref{Rn3}) provides an explicit expression for the kernels $\cK_n$, see eq.(\ref{Kn2}). Next, we can now compute the sum over $n$ in eq.(\ref{G2}), which eventually provides an explicit expression for $B(\br,\sigma)$ using eq.(\ref{G0}). Thus, a simple calculation gives:
\beqa
\lefteqn{ B(\bk,\bp,\tau,\sigma) = e^{3i (\bk.\bp) e^{-\tau} [ 1-e^{\sigma} ]} F(\bk,\bp,\tau-\sigma) } \nonumber \\ & & - 2 i \frac{\bk}{k^2} . \frac{\pl \delta_D}{\pl \bp}(\bp) \int \d\bp' \d\sigma'\d\sigma_1 \int \frac{\d s}{2\pi i} e^{s(\sigma_1-\sigma')} \theta(\sigma-\sigma_1)  \nonumber \\ & & \times \; e^{3i (\bk.\bp) e^{-\tau} [ 1-e^{\sigma-\sigma_1} ] + 3i (\bk.\bp') e^{-\tau} [ e^{\sigma-\sigma'}-e^{\sigma} ] } \nonumber \\ & &  \times \; e^{\tau+\sigma'-\sigma} \; \frac{s(s-1)}{(s-3)(s+2)} \; F(\bk,\bp',\tau-\sigma) .
\label{B5}
\eeqa
It is convenient to replace the integrations over $\sigma'$ and $\sigma_1$ in eq.(\ref{B5}) by integrals over the conjugate Laplace coordinates $s'$ and $s_1$. To do so, we simply compute the integrals over $\sigma'$ and $\sigma_1$ in eq.(\ref{B5}) in terms of the function $\psi(s,x)$ which we define by:
\beq
\psi(s,x) \equiv \int_0^{\infty} \d\sigma \; e^{- s \sigma + x [1-e^{\sigma}]} .
\label{psi1}
\eeq
In the following the argument $x$ of the function $\psi(s,x)$ will be imaginary. Then, we have the asymptotic behaviour for large positive $s$:
\beq
\Re(x)=0 , \; \Re(s) \rightarrow +\infty : \;\; \psi(s,x) \rightarrow 0
\eeq
since for imaginary $x$ and $\Re(s)>0$ eq.(\ref{psi1}) implies: $|\psi(s,x)| \leq 1/\Re(s)$. The integral representation (\ref{psi1}) only applies to $\Re(s)>0$ however the function $\psi$ can be continued to $\Re(s) \leq 0$. Indeed, from eq.(\ref{psi1}) one can check that $\psi(s,x)$ can be expressed in terms of the incomplete Gamma function $\Gamma$ or the confluent hypergeometric function $\psi_K$ (here we add a subscript ``K'' to distinguish Kummer's function $\psi_K$ from $\psi$ defined in eq.(\ref{psi1})) by (e.g., \cite{Bat1}):
\beq
\psi(s,x) = e^x x^s \Gamma[-s,x] = \psi_K(1,1-s;x) .
\label{psiK1}
\eeq
This implies that $\psi(s,x)$ is an entire function of $s$. In particular, one obtains from eq.(\ref{psiK1}) the asymptotic behaviour for large negative $s$ at fixed $x$ (see \cite{Bat1}, vol.I,\S 6.13.2.(6)):
\beq
\Re(s) \rightarrow -\infty : \;\; \psi(s,x) \sim x^s e^{s-(s+1/2) \ln(-s)}
\eeq
which shows that $\psi(s,x)$ diverges for $\Re(s) \rightarrow -\infty$. On the other hand, from eq.(\ref{psi1}) we see that the function $\psi(s,x)$ also satisfies the relation (inverse Laplace transform):
\beq
e^{- s_1 \sigma + x [1-e^{\sigma}]} = \inta \frac{\d s}{2\pi i} \; e^{s \sigma} \; \psi(s+s_1,x) .
\label{psi2}
\eeq
Then, we can write eq.(\ref{B5}) as:
\beqa
\lefteqn{ B(\bk,\bp,\tau,\sigma) =  \int \frac{\d s_1}{2\pi i} e^{s_1 \sigma} \psi(s_1,3i (\bk.\bp) e^{-\tau}) \biggl \lbrace F(\bk,\bp,\tau-\sigma) } \nonumber \\ & & - 2i \; e^{\tau} \frac{\bk}{k^2} . \frac{\pl \delta_D}{\pl \bp}(\bp) \int \frac{\d s_2 \d s_3}{(2\pi i)^2} e^{(s_2+s_3)\sigma} \frac{s_1(s_1-1)}{(s_1-3)(s_1+2)}  \nonumber \\ & & \times \; \frac{1}{s_1+s_2-1} \int \d\bp' \; \psi(s_2,-3i (\bk.\bp') e^{-\tau}) \nonumber \\ & & \times \; \psi(s_3+1,3i (\bk.\bp') e^{-\tau}) \; F(\bk,\bp',\tau-\sigma) \biggl \rbrace .
\label{B6}
\eeqa
Using the definition (\ref{Om1}) this result provides the operator $\cOm$ through the formal integral:
\beq
(\cOm . F)(\br) = \int_0^{\infty} \d\sigma \; B(\br,\sigma) .
\label{OmB1}
\eeq

\section{Calculation of the kernel $\tK(\br,\br_1,\br_2)$}
\label{Calculation of the kernel}

From eq.(\ref{B6}) and the expression (\ref{OmB1}) for the operator $\cOm$ we can derive the explicit form of the kernel $\tK(\br;\br_1,\br_2)$ defined in eq.(\ref{tK1}), using eq.(\ref{K1}). First, we write the factor $\delta_D(\tau_1-\tau) \delta_D(\tau_2-\tau)$ which appears in eq.(\ref{K1}) as $\delta_D(\tau_1-\tau) \delta_D(\tau_2-\tau_1)$. Then, using eq.(\ref{B6}) we see that $B(\br,\sigma) =  e^{-\sigma \cO} . K$ exhibits a factor $\delta_D(\tau_1-\tau+\sigma)$. Thus, the integration on $\sigma$ over the range $[0,\infty[$ is now straightforward and it yields a Heaviside factor $\theta(\tau-\tau_1)$. Next, a simple calculation yields for $\tK(\br;\br_1,\br_2)$:
\beqa
\lefteqn{ \tK = 2i \; e^{\tau_1} \; \theta(\tau-\tau_1) \; \delta_D(\tau_2-\tau_1) \; \delta_D(\bk_1+\bk_2-\bk) } \nonumber \\ & & \times \int \frac{\d s_1}{2\pi i} e^{s_1(\tau-\tau_1)} \psi(s_1,3i (\bk.\bp) e^{-\tau}) \biggl \lbrace \frac{\bk_1}{k_1^2} . \frac{\pl \delta_D}{\pl \bp}(\bp_2-\bp) \nonumber \\ & &  \hspace{0.2cm} - \; 6 \; \frac{\bk_1 . \bk}{k_1^2} \; \frac{\bk}{k^2} . \frac{\pl \delta_D}{\pl \bp}(\bp) \int \frac{\d s_2 \d s_3}{(2\pi i)^2} \; e^{(s_2+s_3)(\tau-\tau_1)} \nonumber \\ & & \hspace{0.5cm} \times \; \frac{s_1(s_1-1)}{(s_1-3)(s_1+2)(s_1+s_2)(s_1+s_2-1)} \nonumber \\ & & \hspace{0.5cm} \times \; \psi(s_2,-3i (\bk.\bp_2) e^{-\tau}) \psi(s_3,3i (\bk.\bp_2) e^{-\tau}) \biggl \rbrace
\label{tK2}
\eeqa
where the integration paths for the variables $s_1$ and $s_2$ obey: $\Re(s_1)>3$ and $\Re(s_2)>0$. To get eq.(\ref{tK2}) we used the following properties of $\psi(s,x)$ which are obvious from eq.(\ref{psi1}):
\beq
\psi(s,0) = \frac{1}{s} , \hspace{0.2cm} \frac{\pl \psi}{\pl x}(s,x) = \psi(s,x) -  \psi(s-1,x) .
\label{psi3}
\eeq
Note the factor $\theta(\tau-\tau_1)$ in eq.(\ref{tK2}) which explicitly shows that the expansion (\ref{rec2}) satisfies causality requirements. This agrees with the discussion in Sect.\ref{Perturbative expansion}.

Finally, as discussed in Sect.\ref{Perturbative expansion} we need to check that the kernel $\tK(\br;\br_1,\br_2)$ obtained in eq.(\ref{tK2}) satisfies eq.(\ref{tK0}). To do so, it is convenient to first integrate over $s_1$ in eq.(\ref{tK2}). The integration of the first term is straightforward, using the inverse Laplace transform (\ref{psi2}). In order to integrate the second term over $s_1$ we use the identity (\ref{Dirac2}). This allows us to express this factor in terms of $\psi(s_1,0)=1/s_1$ and $\psi(s_1-1,0)=1/(s_1-1)$, so that the integration over $s_1$ becomes quite simple. In particular, the contribution from the poles $s_1=-s_2$ and $s_1=1-s_2$ vanishes when we take the integration path over $s_2$ to be $\Re(s_2)>3$. Then, we can write the kernel $\tK$ as:
\beqa
\lefteqn{ \tK = 2i \; e^{\tau_1} \; \delta_D(\tau_2-\tau_1) \; \delta_D(\bk_1+\bk_2-\bk) } \nonumber \\ & & \times \biggl \lbrace \theta(\tau-\tau_1) \; \frac{\bk_1}{k_1^2} . \frac{\pl \delta_D}{\pl \bp}(\bp_2-\bp) \; e^{3i (\bk.\bp) [ e^{-\tau} -e^{-\tau_1} ]} \nonumber \\ & & - \; \frac{6}{5} \; \frac{\bk_1 . \bk}{k_1^2} \inta \frac{\d s_2 \d s_3}{(2\pi i)^2} \; e^{(s_2+s_3)(\tau-\tau_1)} \nonumber \\ & & \times \; \psi\left(s_2,-3i (\bk.\bp_2) e^{-\tau}\right) \; \psi\left(s_3,3i (\bk.\bp_2) e^{-\tau}\right) \nonumber \\ & & \times \biggl [ \frac{\bk}{k^2} . \frac{\pl \delta_D}{\pl \bp}(\bp) \left( \frac{2 \; e^{3(\tau-\tau_1)}}{(s_2+2)(s_2+3)} + \frac{3 \; e^{-2(\tau-\tau_1)}}{(s_2-2)(s_2-3)} \right) \nonumber \\ & & + 3 i e^{-\tau} \delta_D(\bp) \left( \frac{e^{3(\tau-\tau_1)}}{(s_2+2)(s_2+3)} - \frac{e^{-2(\tau-\tau_1)}}{(s_2-2)(s_2-3)} \right) \biggl ] \biggl \rbrace \nonumber \\
\label{tK3}
\eeqa
where the integration path obeys $\Re(s_2)>3$. Note that we removed the Heaviside factor $\theta(\tau-\tau_1)$ from the second term in eq.(\ref{tK3}) because the integrals over $s_2$ and $s_3$ vanish for $(\tau-\tau_1)<0$ (since for $(\tau-\tau_1)<0$ we can push the integration path to $\Re(s) \rightarrow +\infty$). Next, we can apply the operator $\cO$ defined in eq.(\ref{O1}) onto the kernel $\tK$ written in eq.(\ref{tK3}). After a tedious calculation we finally obtain eq.(\ref{tK0}). The only technical trick in this calculation is to use eq.(\ref{psi3}) as well as the property:
\beq
s \; \psi(s,x) = 1 - x \; \psi(s-1,x) .
\label{psi4}
\eeq
This identity is obtained by multiplying eq.(\ref{psi1}) by $s$ and integrating by parts. This explicit check of eq.(\ref{tK0}) fully justifies the procedure developed in Sect.\ref{Building a perturbative expansion}.

\end{document}